\documentclass[pre,aps,onecolumn,amsmath,showpacs,floatfix]{revtex4}
\usepackage{graphicx}

\usepackage{amssymb}
\usepackage{natbib}
\usepackage{dcolumn}
\usepackage{verbatim}

\begin{document}

\title{Investigation of resonant and transient phenomena in Josephson
junction flux qubits}
\author{Jeffrey E.~Marchese}
\affiliation{Department of Applied Science,$\;$University of California, Davis,
California 95616}
\author{Matteo Cirillo}
\affiliation{Dipartimento di Fisica and MINAS-Lab, Universit\`{a} di Roma
\textquotedblleft Tor Vergata", I-00173 Roma, Italy}
\author{Niels Gr{\o }nbech-Jensen}
\affiliation{$\;$Department of Applied Science,$\;$University of California, Davis,
California 95616$\;$}
\date{\today}

\begin{abstract}
We present an analytical and computational study of resonances and transient
responses in a classical Josephson junction system. A theoretical basis for
resonances in a superconducting loop with three junctions is presented,
outlining both the direct relationship between the dynamics of single- and
multi-junction systems, and the direct relationships between observations of
the classical counterparts to Rabi oscillations, Ramsey fringes, and spin
echo oscillations in this class of systems. We show simulation data along
with analytical analyses of the classical model, and the results are related
to previously reported experiments conducted on three junction loops. We
further investigate the effect of off-resonant microwave perturbations to,
e.g., the Rabi-type response of the Josephson system, and we relate this
response back to the nonlinear and multi-valued resonance behavior
previously reported for a single Josephson junction. The close relationships
between single and multi-junction behavior demonstrates the underlying
dynamical mechanism for a whole class of classical counterparts to expected
quantum mechanical observations in a variety of systems; namely the resonant
and transient behavior of a particle in an anharmonic potential well with
subsequent escape.
\end{abstract}

\maketitle


\section{Introduction}

Since the mid 1980's several experiments have sought to establish that
Josephson systems show evidence of quantum interference of macroscopically
distinct states. Various configurations have been utilized to elicit the
phenomenon, including current-biased single-junction circuits \cite%
{silvestrini.book1,silvestrini:mqt,martinis:rabi,Simmonds,Wallraff_03},
flux-biased two-junction SQUIDs \cite{claudon.buisson:rabi}, charge-biased
two-junction SQUIDs \cite{wallraff.devoret:ramsey}, flux-biased
three-junction SQUIDs \cite{vanderwal:superposition,
chiorescu.nakamura:rabi,Ilichev:rabi,kutsuzawa.takayanagi:ramsey,plourde.clarke:ramsey,saito.takayanagi:osc}%
, charge-biased three-junction SQUIDs \cite{vion.esteve.devoret:sciramsey},
and hybrid systems which often include inductor-based systems to create
multiple potential wells \cite%
{friedman:superposition,nakamura:cooper,pashkin.nakamura:rabi,koch:ramsey,mooij:spin-echo}%
. The phenomena observed to date include multi-peaked probability
distributions in bias current, Rabi oscillations\cite{Rabi}, Ramsey fringes%
\cite{Ramsey}, spin-echo oscillations\cite{Vion_03}, and quantum tomography%
\cite{Martinis:tomo}.

In several recent investigations the authors of the present paper have
contributed to an alternative interpretation of the basic types of
experiments on Josephson junctions. This interpretation relies on the well
established phenomenology of the Resistively and Capacitively Shunted
Junction (RCSJ) model \cite{vanduzer.barone} and on nonlinear dynamics
related to it. Multi-peaked resonances were addressed in Refs.[25-27] while
References~[28,29] developed analytical expressions and presented computer
simulations which described the relationships between classical Rabi-type
oscillations, driving amplitude and frequency, and systematic dissipation.
The related phenomena of Ramsey-fringes and spin-echo oscillations are
depicted and described from a classical point of view in Refs.~[30,31].
While this exemplifies the phenomena through current-biased single-junction
circuits, we submit that a wide class of Josephson systems exhibits same
behavior when a system variable is trapped in a potential well \cite%
{Jeff_Thesis}. This presentation will amplify this point.

A few examples of multi-junction systems for which direct comparisons
between classical theory and experimental results have been made in the low
temperature regime, where quantum phenomena should be observed, include
multi-peaked resonances in Josephson interferometers, made of two junctions
connected in parallel \cite{physicac}, and a more complex Josephson
interferometer-based device\cite{catastrophe} containing a double-SQUID, in
which a butterfly catastrophe was observed at 10 mK. In both cases, the
standard classical RCSJ model provided sufficient detail to understand the
experimental observations.

The observations from interferometer systems described in the last paragraph
encouraged us \cite{Jeff_Thesis} to extend the classical ideas and
implementations developed for the single junctions and two-junction
interferometers to the more complex system of the flux-biased three-junction
loop \cite%
{Ilichev:rabi,kutsuzawa.takayanagi:ramsey,plourde.clarke:ramsey,mooij:qubit}%
. It is the plan of the present paper to report on our approach to this
problem. In this approach the analysis of the most important features of
three junctions loop under the influence of pulsed microwaves was preluded
by a study of analogous single junction phenomenology.

In the next section we derive the basic equations for the three junction
loop and present an analytical description of resonances in the system. In
sect. 3 we demonstrate computer simulations which map the resonances as a
function of the biases (the magnetic flux in the loop). Simulations are
shown which depict the Rabi-type oscillations, Ramsey-type fringes and
spin-echo-type oscillations in the three-junction loop. Also, off-resonant
behavior in the three-junction loop is explored.

Our simulations and analytical treatment allow us to interpret the essential
experimental features reported for Josephson circuits; the analysis shall
all be based on the RCSJ modelling of Josephon circuits\cite{vanduzer.barone}
who took origin from early remarks on the classical (i.e. , continuous)
nature of Josephson phase and flux variables. In this context we also refer
to the subjects of our presentation as "classical" analyses and modelling.

\section{Three-junction loop RCSJ model}

\subsection{Single-junction theory}

Figure~1(a) depicts the single Josephson junction RCSJ
model. Such a physical system is also referred to as Josephson phase qubit.
The normalized classical equation can be written\cite{vanduzer.barone} 
\begin{equation}
\ddot{\varphi}+\alpha \dot{\varphi}+\sin \varphi =\eta +\varepsilon _{s}\sin
(\omega _{s}t+\theta _{s})+\varepsilon _{p}(t)+n(t),
\end{equation}%
where $\varphi $ is the difference between the phases of the quantum
mechanical wave functions defining the junction, $\eta $ represents the dc
bias current, and $\varepsilon _{s}(t)$ and $\omega _{s}$ represent
microwave current amplitude and frequency, respectively. All currents are
normalized to the critical current $I_{c}$, and time is measured in units of
the inverse plasma frequency, ${\omega _{0}}^{-1}$, where ${\omega _{0}}%
^{2}=2eI_{c}/\hbar C=2\pi I_{c}/\Phi _{0}C$, $C$ being the capacitance of
the junction and $\Phi _{0}=h/2e$ the flux quantum. Tunneling of
quasiparticles is represented by the dissipative term, where $\alpha =\hbar
\omega _{0}/2eRI_{c}$ is given by the shunt resistance $R$, and the
accompanying thermal fluctuations are time-correlated by the normalized
fluctuation-dissipation relationship\cite{Parisi_88} 
\begin{eqnarray}
\left\langle n(t)\right\rangle &=&0 \\
\left\langle n(t)n(t^{\prime })\right\rangle &=&2\alpha {\frac{k_{B}T}{H_{J}}%
}\delta (t-t^{\prime })=2\alpha \Theta \delta (t-t^{\prime }),
\end{eqnarray}%
$T$ being the temperature and $H_{J}$ is the characteristic Josephson energy 
$H_{J}=I_{c}\hbar /2e$.

The normalized energy, ($H-H_{0}$) is defined by 
\begin{eqnarray}
H &=&\frac{1}{2}\dot{\varphi}^{2}+1-\cos \varphi -\eta \varphi  \label{eq:H}
\\
H_{0} &=&1-\sqrt{1-\eta ^{2}}-\eta \sin ^{-1}\eta \;,  \label{eq:H_0}
\end{eqnarray}

where $H_{0}$ is the energy at the minimum of a potential well. In
Refs.[28,29] a classical analysis is presented for the modulated frequency,
which is small when compared to the driving signal $\omega _{s}$, and
referred to as the Rabi-type oscillation frequency $\Omega _{R}$ due to its
direct relationship with experimentally reported Rabi oscillations. Thus,
the classical analog to quantum mechanical Rabi oscillations has been
denoted Rabi-type oscillations [28].
\begin{figure}
[pt]
\begin{center}
\includegraphics[
trim=0.000000in 0.103025in 0.000000in 0.068900in,
height=3.50000in,
width=2.510500in
]%
{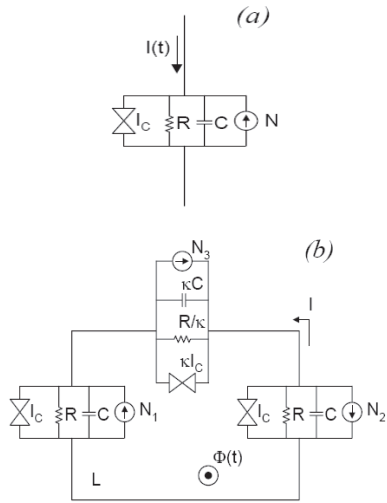}%
\caption{Circuit diagram for (a) the single junction model and (b) the three
junction loop. In (b) the smaller junction is indicated by the factor $%
\protect\kappa $ appklied to the capacitance, resistance , and critical
current.
.}%
\label{Fig.1}%
\end{center}
\end{figure}

\subsection{\protect\bigskip Three junction loop theory}

Figure~1b depicts the three-junction loop which will be the
subject of the analysis of this paper. We concentrate our study on this
system which has been the core of several experiments and theoretical
analyses\cite{mooij:qubit}. It consists of a superconducting loop containing
two identical junctions and one with smaller capacity and critical current.
The normalized classical equations can be written \cite%
{geigenmuller:tjsquid,mooij:qubit},

\begin{eqnarray}
\ddot{\varphi _{1}}+\alpha \dot{\varphi _{1}}+\sin \varphi _{1} &=&{\frac{-1%
}{\beta _{L}}}\left[ 2\pi M(t)+\varphi _{1}+\varphi _{2}+\varphi _{3}\right]
+n_{1}(t)  \nonumber  \label{eq:Main} \\
\ddot{\varphi _{2}}+\alpha \dot{\varphi _{2}}+\sin \varphi _{2} &=&{\frac{-1%
}{\beta _{L}}}\left[ 2\pi M(t)+\varphi _{1}+\varphi _{2}+\varphi _{3}\right]
+n_{2}(t) \\
\ddot{\varphi _{3}}+\alpha \dot{\varphi _{3}}+\sin \varphi _{3} &=&{\frac{-1%
}{{\kappa \beta _{L}}}}\left[ 2\pi M(t)+\varphi _{1}+\varphi _{2}+\varphi
_{3}\right] +n_{3}(t).  \nonumber
\end{eqnarray}%
where $\varphi _{i}$ is the difference between the phases of the quantum
mechanical wave functions defining the $i$th junction, $\kappa $ is the
scale factor relating the one smaller junction, represented by $\varphi _{3}$%
, to the other two larger junctions, $\beta _{L}$ is the normalized loop
inductance $\beta _{L}=2\pi LI_{c}/\Phi _{0}$. $M(t)$ is a variable which
summes up the external magnetic fluxes to which the system can be exposed 
\begin{equation}
M(t)=M_{dc}+\varepsilon _{s}(t)\sin (\omega _{s}t+\theta _{s})+\varepsilon
_{p}(t),
\end{equation}%
where $M_{dc}$ represents the flux of a time-independent field, $\varepsilon
_{s}(t)$ and $\omega _{s}$ represent the ac flux (signal) amplitude and
frequency, respectively while a flux pulse for probing the state of the
system is here taken into account by $\varepsilon _{p}(t)$. A general
analysis of the above system of equations for arbitrary values of the
parameters would surely lead to rather intriguing dynamics and complex
oscillations. In the analysis that we will present herein we will focus on
the limits that are suggested by the experimental reality. A very indicative
example of the effects that a finite inductance can have on Josephson flux
qubit was reported in ref. 34.

We write the Hamiltonian for the undamped ($\alpha =0$) system as 
\begin{eqnarray}
H &=&H_{k}+H_{p}  \label{eq:re-pe} \\
H_{k} &=&\frac{1}{2}(\dot{\varphi}_{1}^{2}+\dot{\varphi}_{2}^{2}+\kappa \dot{%
\varphi}_{3}^{2}) \\
H_{p} &=&2-\cos \varphi _{1}-\cos \varphi _{2}+\kappa (1-\cos \varphi _{3})+%
\frac{1}{2\beta _{L}}(\varphi _{1}+\varphi _{2}+\varphi _{3}+2\pi M)^{2}.
\end{eqnarray}%
For small $\beta _{L}$ in Eq.~(\ref{eq:re-pe}), we can see that energy from
the loop current (last term) dominates the sum of the phases. We introduce
now a transformation which allows combining the phases $\varphi _{1},\varphi
_{2},$and $\varphi _{3}$ \ while maintaining the kinetic part of the
Hamiltonian still separable in three distinct variables ($\psi _{1},\psi
_{2},$ and $\psi _{3}$)

\[
\left. 
\begin{array}{lcl}
\psi _{1} & = & \varphi _{1}+\varphi _{2}+\varphi _{3} \\ 
\psi _{2} & = & \varphi _{1}-\varphi _{2} \\ 
\psi _{3} & = & -\frac{1}{1+2\kappa }(\varphi _{1}+\varphi _{2})+\frac{%
2\kappa }{1+2\kappa }\varphi _{3}%
\end{array}%
\right\} \Leftrightarrow \left\{ 
\begin{array}{lcl}
\varphi _{1} & = & \frac{\kappa }{1+2\kappa }\psi _{1}+\frac{1}{2}\psi _{2}-%
\frac{1}{2}\psi _{3} \\ 
\varphi _{2} & = & \frac{\kappa }{1+2\kappa }\psi _{1}-\frac{1}{2}\psi _{2}-%
\frac{1}{2}\psi _{3} \\ 
\varphi _{3} & = & \frac{1}{1+2\kappa }\psi _{1}+\psi _{3}%
\end{array}%
\right. 
\]%
The Hamiltonian now becomes 
\begin{eqnarray}
H_{k} &=&\frac{1}{2}(\kappa \dot{\psi}_{1}^{2}+\frac{1}{2}\dot{\psi}_{2}^{2}+%
\frac{1}{2}(1+2\kappa )\dot{\psi}_{3}^{2})  \nonumber  \label{eq:energy} \\
H_{p} &=&2-2\cos \frac{\psi _{2}}{2}\cos \left( \frac{\kappa }{1+2\kappa }%
\psi _{1}-\frac{1}{2}\psi _{3}\right) \\
&+&\kappa \left[ 1-\cos \left( \frac{1}{1+2\kappa }\psi _{1}+\psi
_{3}\right) \right] +\frac{1}{2\beta _{L}}(\psi _{1}+2\pi M)^{2}.  \nonumber
\end{eqnarray}

Under the condition $\beta_L \ll 1$, this system can be simplified to a
single degree of freedom by examining the constraints which frame $\psi_1 
\text{ and } \psi_2$. First, it can be seen from Eq.~(\ref{eq:energy}) that
potential energy considerations prevent the deviation of $\psi_1$ far from $%
-2\pi M$.

A second constraint, on $\psi _{2}$, is obtained by considering the equation
of motion 
\[
\ddot{\psi}_{2}+\alpha \dot{\psi}_{2}+2\cos \left( \frac{\kappa }{1+2\kappa }%
\psi _{1}-\frac{1}{2}\psi _{3}\right) \sin \frac{\psi _{2}}{2}%
=n_{1}(t)-n_{2}(t). 
\]%
From this it can seen that the static equilibrium value of $\psi _{2}$ is
given by

\begin{equation}
\cos \frac{\psi _{2}}{2}=\left\{ 
\begin{array}{rcl}
1 & \mathrm{for} & \cos (\frac{\kappa }{1+2\kappa }\psi _{1}-\frac{1}{2}\psi
_{3})>0 \\ 
-1 & \mathrm{for} & \cos (\frac{\kappa }{1+2\kappa }\psi _{1}-\frac{1}{2}%
\psi _{3})<0%
\end{array}%
\right.  \label{ref:eq:costwo}
\end{equation}%
Note that this is a \emph{weak} constraint as it does not require much
energy to break the equilibrium. Equation~(\ref{eq:energy}) can now be
written 
\begin{eqnarray}
H_{p} &=&2\mp 2\cos (\frac{\kappa }{1+2\kappa }\psi _{1}-\frac{1}{2}\psi
_{3})+\kappa (1-\cos (\frac{1}{1+2\kappa }\psi _{1}+\psi _{3}))+\frac{1}{%
2\beta _{L}}(\psi _{1}+2\pi M)^{2},  \nonumber \\
&&
\end{eqnarray}%
where $\pm $ refers to the optimized choice based on Eq.~(12). The
constraint on $\psi _{1}$ can be determined by minimizing $H_{p}$ with
respect to $\psi _{1}$ 
\begin{eqnarray}
\frac{\partial H_{p}}{\partial \psi _{1}} &=&\frac{\pm 2\kappa }{1+2\kappa }%
\sin \left( \frac{\kappa }{1+2\kappa }\psi _{1}-\frac{1}{2}\psi _{3}\right) 
\nonumber \\
&\phantom{.}&+\frac{\kappa }{1+2\kappa }\sin \left( \frac{1}{1+2\kappa }\psi
_{1}+\psi _{3}\right) +\frac{1}{\beta _{L}}\left( \psi _{1}+2\pi M\right)
\;=\;0  \nonumber \\
\Rightarrow \;\;\psi _{1} &=&-2\pi M-\beta _{L}\delta _{1}\;  \nonumber \\
&\approx &\;-2\pi M-\beta _{L}\frac{\kappa }{1+2\kappa }\left[ \mp 2\sin
\left( \frac{2\pi M\,\kappa }{1+2\kappa }+\frac{1}{2}\psi _{3}\right) +\sin
\left( \psi _{3}-\frac{2\pi M}{1+2\kappa }\right) \right] ,  \nonumber
\end{eqnarray}%
which is correct to first order in $\beta _{L}$.

The potential energy can now be aproximated by a function of $\psi _{3}$
only 
\[
H_{p}=2\mp 2\cos \left( \frac{2\pi M\,\kappa }{1+2\kappa }+\frac{1}{2}\psi
_{3}\right) +\kappa -\kappa \cos \left( \psi _{3}-\frac{2\pi M}{1+2\kappa }%
\right) -\frac{\beta _{L}}{2}\delta _{1}^{2}, 
\]%
where the choice between $\mp $ should be such that the energy is minimized.%
\begin{figure}
[pt]
\begin{center}
\includegraphics[
trim=0.000000in 0.103025in 0.000000in 0.068900in,
height=2.50000in,
width=4.000000in
]%
{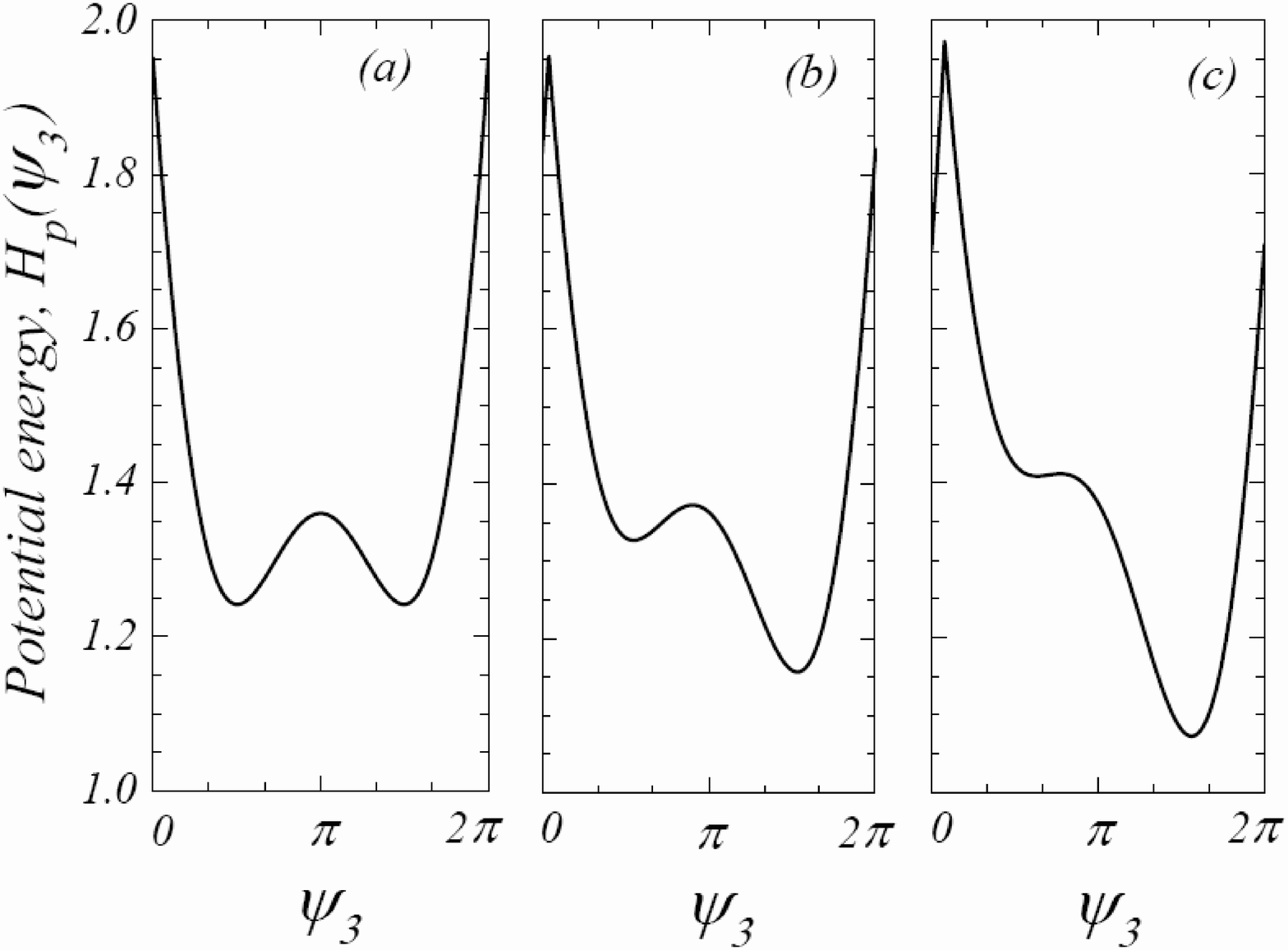}%
\caption{
Three-junction loop Potential
Energy in terms of the reduced variable $\protect\psi _{3}$. Each view
represents a different value of $M_{dc}$: (a) $M_{dc}=0.5$, (b) $M_{dc}=0.52$%
, and (c) $M_{dc}=0.542$. Other parameters are $\protect\kappa =0.68$ and $%
\protect\beta =0.09559$.}%
\label{Fig.2}%
\end{center}
\end{figure}

Figure~2 depicts this single-degree-of-freedom
representation of the potential energy. Note that while, in general, this
potential consists of pairs of wells separated by higher energy
\textquotedblleft cusps", for purposes of this work we focus on a single
well-pair, as the energies are kept below these cusp

Calculating the linear resonance frequency for a given well, the equation of
motion for the friction-less and constant-$M$ system is 
\[
\frac{1}{2}(1+2\kappa )\ddot{\psi}_{3}+\frac{\partial H_{p}}{\partial \psi
_{3}}=0, 
\]%
such that the fixed point $\psi _{3}^{\ast }$ is given by 
\[
\left. \frac{\partial H_{p}}{\partial \psi _{3}}\right\vert _{\psi _{3}=\psi
_{3}^{\ast }}=0 
\]%
and for small oscillations $\delta $ around the fixed point $\psi _{3}=\psi
_{3}^{\ast }+\delta $ provide the linear resonance frequency $\omega _{l}$ 
\begin{eqnarray}
&&\frac{1}{2}(1+2\kappa )\ddot{\delta}+\left. \frac{\partial ^{2}H_{p}}{%
\partial \psi _{3}^{2}}\right\vert _{\psi _{3}^{\ast }}\;\delta
\;\;=\;\;0\;\;\Rightarrow  \label{eq:LinResFreq} \\
&&\omega _{l}\;\;=\;\;\sqrt{2\left. \frac{\partial ^{2}H_{p}}{\partial \psi
_{3}^{2}}\right\vert _{\psi _{3}^{\ast }}/(1+2\kappa )}\;.
\end{eqnarray}%
In the next section we will use this result to characterize the switching
statistics of microwave induced excitations in the classical model of the
three-junction superconducting loop.

\section{Simulation results}

Our system is periodic under the application of the extrenal magnetic flux.
In order to best describe the experimental reality the temperatures that we
consider are in the tens of millikelvin range: we point out, however, that
the kind of phenomena like those that we shall describe herein have been
characterized in a temperature range spanning from few tens of mK up to 800
mK\cite{Jeff_Thesis}; we believe that most of the phenomena herein described
would display quite similar dependencies on temperature over the same range.

To verify equation~(\ref{eq:LinResFreq}), we conducted computer simulations
of pulsed microwave spectroscopy for the three-junction loop described by
equations (\ref{eq:Main}); naturally, the simulations are performed on the
entire system of equations regardless of the analytical approximations.
Microwave spectroscopy, combined with thermal escape measurements has turned
out to be a reliable tool for investing resonances[25-27]; here we try to
follow real experimental recipes\cite{physicac}; we must bear in mind that
for the thermal escape we simulate the whole system of Eq.6 and that
switches to voltage states are just detected across a parallel connection of
two junctions which forces us to attribute a 50\% equal probability of
switching to the two junctions. The results of our tracing the resonance
through thermal escape and microvave pulses are shown in Fig.~3.
Each open circle represents the minimum flux pulse $%
\varepsilon _{p}$ required to obtain a statistical response of $50\%$ escape
probability. Each trial in the simulation is conducted in the following
manner: with initial conditions of $\psi _{3}$ in the minimum energy
configuration and $\dot{\psi _{3}}=0$, the system is driven at frequency $%
\omega _{s}$ and amplitude $\varepsilon _{s}$ for a period of $(2/\alpha )$.
At the end of the driving phase and after a slight pause (50 time units), a
pulse of magnitude $\varepsilon _{p}$ is applied (in similar fashion to
reference~[30]) and the subsequent escape or non-escape to the alternate
well is recorded. Statistics are then gathered to determine $\varepsilon
_{p}^{50\%}$ for varying frequency. A resonance is determined for each value
of $M_{dc}$, as indicated by a minimum value of $\varepsilon _{p}^{50\%}$ as
a function of frequency. In the inset of Fig. 3 we show the resonance effect
in terms of switching probability, exemplified 
\begin{figure}
[pt]
\begin{center}
\includegraphics[
trim=0.000000in 0.103025in 0.000000in 0.068900in,
height=3.50000in,
width=4.000000in
]%
{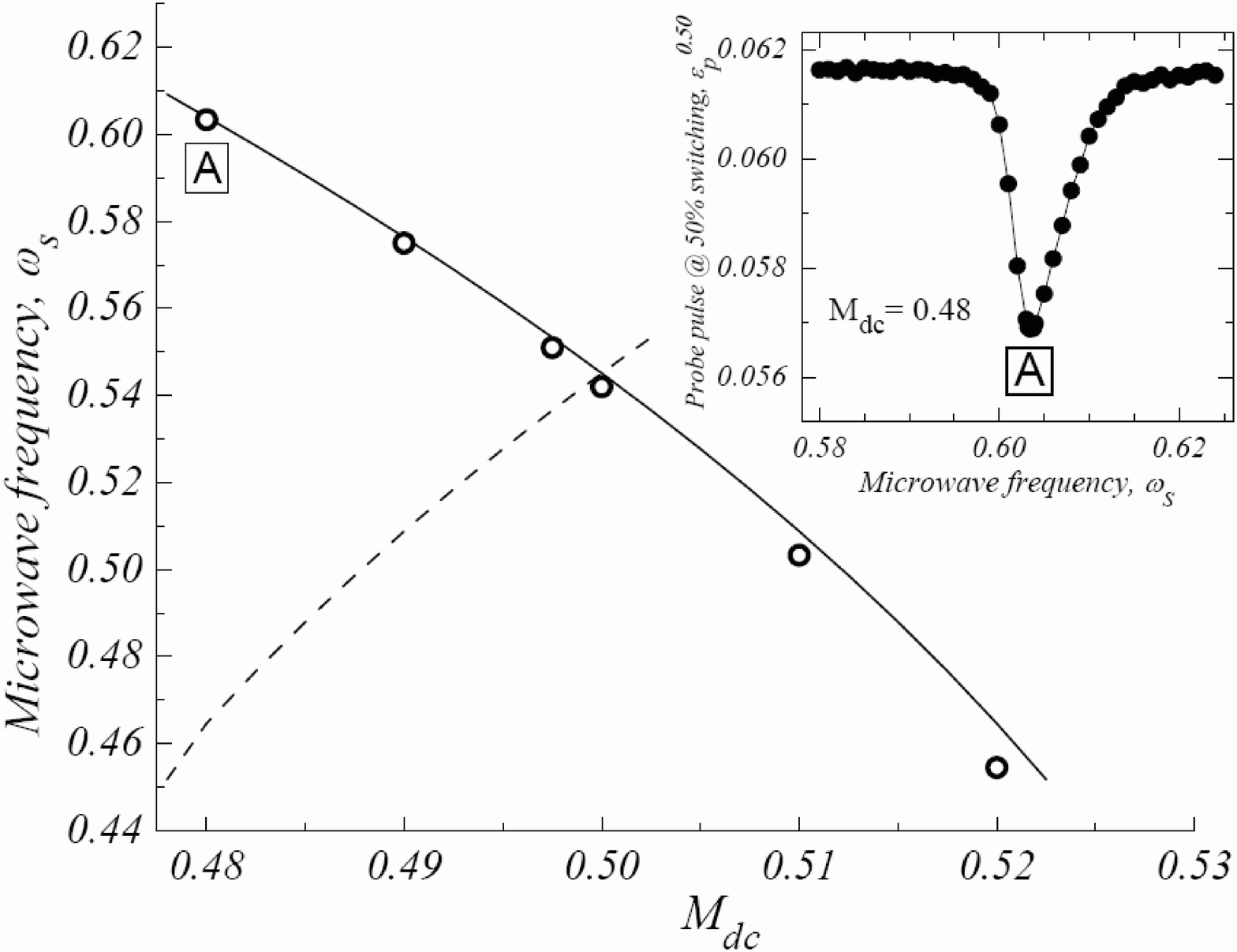}%
\caption{Resonant frequency response in the three-junction loop.
Lines represent the predicted frequency response from Eq.~(\protect\ref%
{eq:LinResFreq}). The open circles indicate the resonances from the results
of our simulations. The inset is a switching distribution indicating the
resonance at the box marked 'A'. This provides an example of the placement
for the open circles in the larger plot. Parameters for the simulation were $%
\protect\beta _{L}=0.09994$, $\protect\kappa =0.68$, $\protect\alpha =4\cdot
10^{-5}$, $\protect\varepsilon _{s}=4\cdot 10^{-4}$, and $\Theta
(T)=9.12\cdot 10^{-3}$. The dashed line indicates the symmetric relationship
in linear resonance frequency. The data presented is based on 2500-7000
escape events.
}%
\label{Fig.3}%
\end{center}
\end{figure}

The solid line in Fig.~3 represents the resonant
frequency response as predicted by Eq.~(\ref{eq:LinResFreq}). The open
circles indicate the simulation results and depict the minimum amplitude
probe pulse $\varepsilon _{p}$ for which the escape rate equals 50\%. The
inset shows the relationship between probe pulse (at 50\% escape
probability) and signal frequency for $M_{dc}=0.48$. The minimum energy
value in each well is marked with a boxed symbol to establish the
correspondence with the potential energy plot. The dashed line indicates the
symmetric relationship in linear resonance frequency (about $M_{dc}=0.5$)
for a well placement convention which is opposite to that of the continuous
curve: in other terms the continuous line we trace the resonance moving the
lower energy well (follow Fig. 2) from left to right while the dashed line
would correspond to tracing the resonance moving the lower energy well from
right to left. The data presented is based on 2500-7000 escape events ; the
simulation results show close agreement with our theory although the
agreement diminishes slightly for increasing values of $M_{dc}$. This can be
understood by noting that as the flux bias is increased, the higher of the
two wells becomes broader, allowing larger oscillations, which corresponds
to a greater degree of anharmonic behavior.

Next we investigated whether the results obtained on the linear resonance
frequency calculated from Eq. \ref{eq:LinResFreq} will generate features
consistent with what one could expect from spectroscopic measurements on
real Josephson junction systems. Thus we tuned the parameters of our
simulations to generate Rabi-type oscillations, Ramsey-type fringes, and
spin-echo-type oscillations\cite{marchese:ramsey_1,marchese:ramsey_2}\cite%
{Jeff_Thesis}. The recipes for generating these oscillations follow closely
the ones described in the previous publications for the single junction
model and a summary of the signalling used for generating Rabi-type
oscillations, Ramsey-type fringes, and spin-echo is shown in 
Figure~4
together with the idealized phase responses in the two relevant cases of
escape and no-escape from the potential well.
\begin{figure}
[pt]
\begin{center}
\includegraphics[
trim=0.000000in 0.000000in 0.000000in 0.000000in,
height=5.00000in,
width=2.000000in
]%
{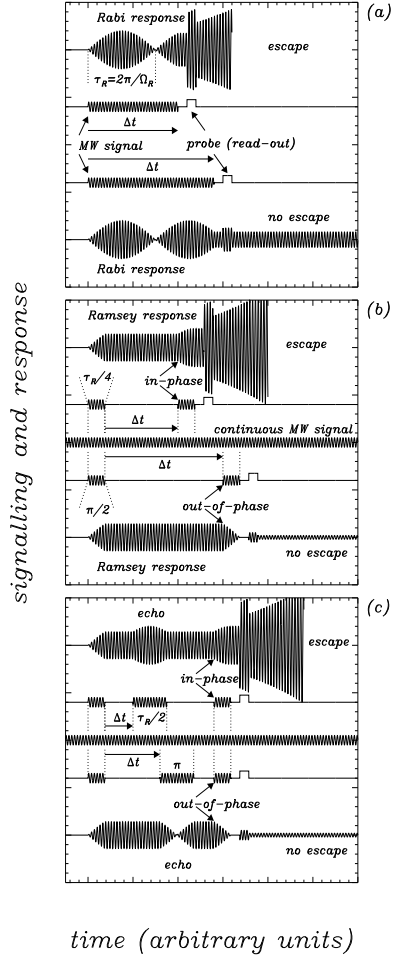}%
\caption{Sketches of idealized microwave perturbations and
corresponding classical system (phase) responses for (a) Rabi oscillations,
(b) Ramsey fringes, and (c) spin echo. Upper part of each plot illustrates a
situation in which the probe (read out) pulse results in an escape from the
potential well, while the lower part illustrates a no-escape case. The
experimentally varied time for observing the oscillations in response is
shown on each plot as $\Delta t$.
}%
\label{Fig.4}%
\end{center}
\end{figure}
Our statistical simulation results are shown in Figs.~5 and %
6 . The normalized slope that we extract from the linear
dependency of the Rabi frequency versus ac amplitude of Fig. 5b is very
close to that of the experimental results reported in Fig. 4b of Ref. [12]
and with Fig. 4b of Ref.~[10]. We conclude that our numerical and analytical
results are quite consistent with the experimental reality. As far as the
Ramsey-type fringes are concerned, if we compare Fig. 5d in this
presentation with the results reported in Fig. 4d \ of Ref.~[12], we also
find very good agreement in the Ramsey-type fringe being represented by a
unity-sloped V-shaped fringe frequency dependence on the microwave detuning
from the oscillator resonance. We clearly see our V-shape following the
dashed lines in Fig. 5d. Thus, just as it has been demonstrated for Rabi
oscillations, the classical system exhibits a signature similar, if not
identical, to what one would expect from quantum theory.

\begin{figure}
[pt]
\begin{center}
\includegraphics[
trim=0.000000in 0.000000in 0.000000in 0.000000in,
height=3.00000in,
width=5.000000in
]%
{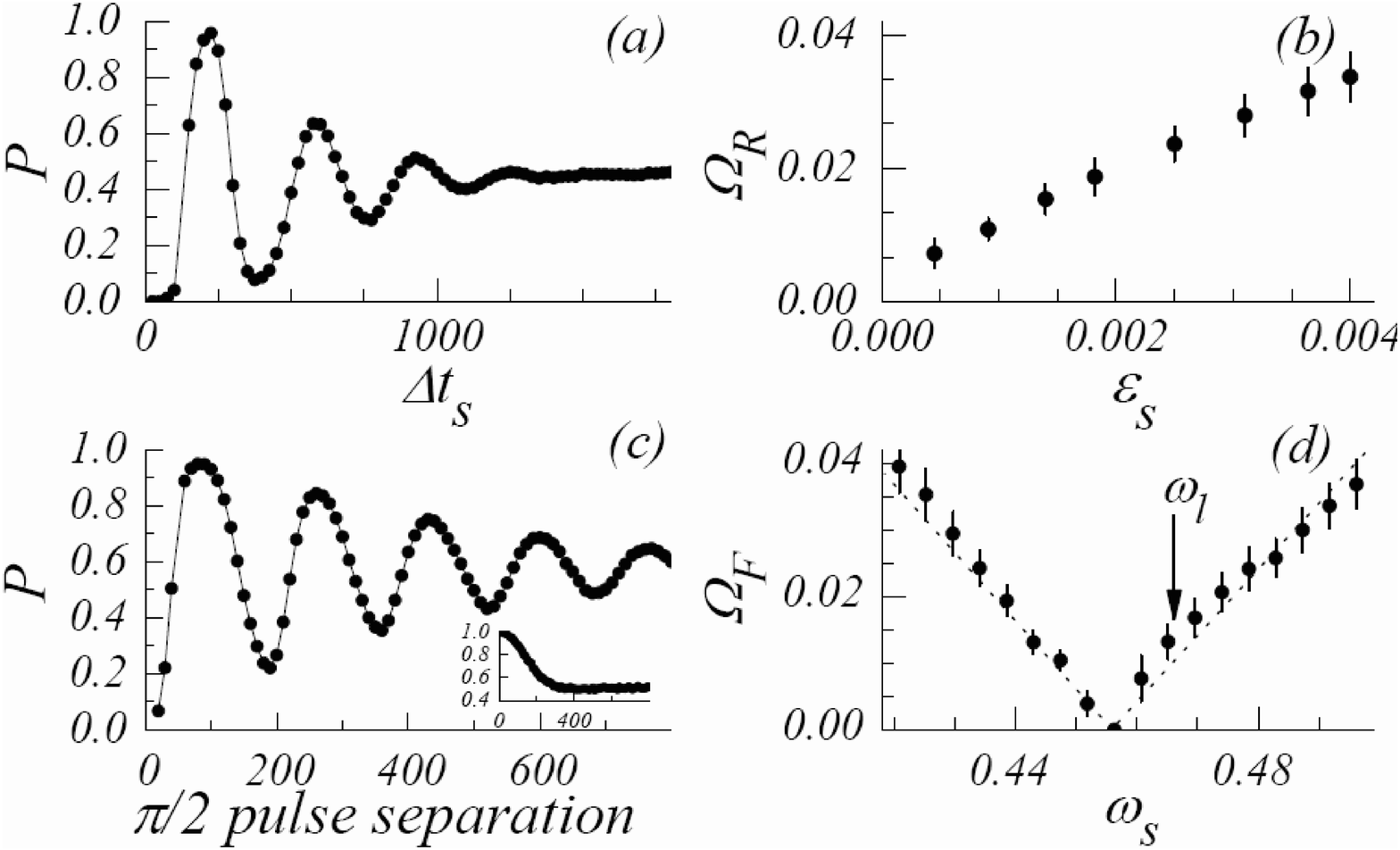}%
\caption{Oscillation phenomena for the
three-junction loop: Rabi-type oscillations and Ramsey-type fringes.
Panel~(a) shows Rabi-type oscillations. Panel~(b) indicates the Rabi-type
frequency as a function of microwave amplitude, with $\protect\varepsilon %
_{s}$ for $\protect\omega _{s}=0.45629\sim \protect\omega _{r}$. Panel~(c)
is the resulting switching distribution for Ramsey-type fringes with $%
\protect\omega _{s}=0.42528$ and $\protect\varepsilon _{p}=0.0195$. The
inset depicts the driving frequency which achieves a fringe frequency of
zero; here $\protect\omega _{s}=0.45629\sim \protect\omega _{r}$, $\protect%
\varepsilon _{p}=0.0148$. Panel~(d) provides the relationship between fringe
frequency and driving frequency. The arrow indicates measurement of $\protect%
\omega _{l}=0.46557$ by direct simulation. Note Eq.~(\protect\ref%
{eq:LinResFreq}) predicts $\protect\omega _{l}=0.46625$. Parameters for all
panels, unless otherwise noted, were $M_{dc}=0.52$, $\protect\beta %
_{L}=0.09559$, $\protect\kappa =0.68$, $\protect\alpha =1.5\cdot 10^{-4}$, $%
\protect\varepsilon _{s}=1.82\cdot 10^{-3}$, $\protect\omega _{s}=0.443$, $%
\protect\varepsilon _{p}=0.0149$, and $\Theta (T)=3\cdot 10^{-3}$;
statistics were gathered for $\sim $ 20 000 escape events.
}%
\label{Fig.5}%
\end{center}
\end{figure}

We note that our analytical approximations explain how the three-loop
potential can be reduced to a single degree of freedom potential, however,
it is clear from the results of the direct integration of eqs. (6) shown in
Fig. 6 that the response of the loop is quite similar to that of a single
junction \cite{marchese:ramsey_1,marchese:ramsey_2}.

The classical result, leading to the Ramsey fringe frequency being identical
to the detuning between the applied microwave frequency $\omega _{s}$ and
the intrinsic resonance $\omega _{r}$ of the oscillator, can be rationalized
by considering the system behavior in the interval between the two $\pi /2$%
-pulses. The oscillator is, at the time of the conclusion of the first $\pi
/2$-pulse, defined by a specific phase-relationship (phase-locking) to the
phase of the $\pi $-pulse. This phase-relationship is free to detune in the
(ballistic) interval $\Delta t$ between the two $\pi /2$-pulses, and the
mutual phase detuning $\theta _{F}$ between the microwave field and the
oscillator is therefore given by 
\[
\theta _{F}=\int_{0}^{\Delta t}(\omega _{s}-\omega _{r}(t))\,dt\;, 
\]%
where $\omega _{r}$ can be a (weak) function of time due to the
anharmonicity of the potential energy; i.e., as the oscillation amplitude
decreases, due to the intrinsic damping in the system, the natural resonance
frequency of the oscillator increases. However, for a lightly damped system,
where $\alpha \Delta t\ll 1$, the above time integration can be simplified,
and the result for the Ramsey fringe frequency $\Omega _{F}$ becomes 
\[
\Omega _{F}=\frac{\theta _{F}}{\Delta t}\;\approx \;\omega _{s}-\omega
_{r}\;, 
\]%
which is exactly the detuning between the applied microwave frequency and
the natural resonance frequency of the oscillator in the phase-locked state.
Thus, the observed Ramsey fringe frequency dependence on the microwave
detuning from resonance is easily understood classically for this system,
and the intrinsic oscillator frequency $\omega _{r}\lesssim \omega _{l}$ is
close to the microwave frequency that results in $\Omega _{F}\approx 0$. We
submit that this detuning is also causing the so-called spin-echo
measurements, which are generated by maintaining a constant temporal
separation between the two $\pi /2$-pulses from Ramsey fringe measurements,
and then inserting a $\pi $-pulse. The final phase-relationship between the
oscillator and the external microwave signal is detected by the second $\pi
/2$-pulse, but this phase-relationship is now subject to the phase-twisting
caused by the inserted $\pi $-pulse, which in turn depends on the detuning
at the time of $\pi $-pulse initiation. Thus, while the Rabi-frequency
determines the magnitude and duration of the microwave pulses used in
Ramsey-fringe experiments, it is the Ramsey-fringe frequency that determines
the echo from the detuning at the time of the initiation of the $\pi $-pulse 
\cite{marchese:ramsey_2}.
\begin{figure}
[pt]
\begin{center}
\includegraphics[
trim=0.000000in 0.000000in 0.000000in 0.000000in,
height=3.00000in,
width=5.000000in
]%
{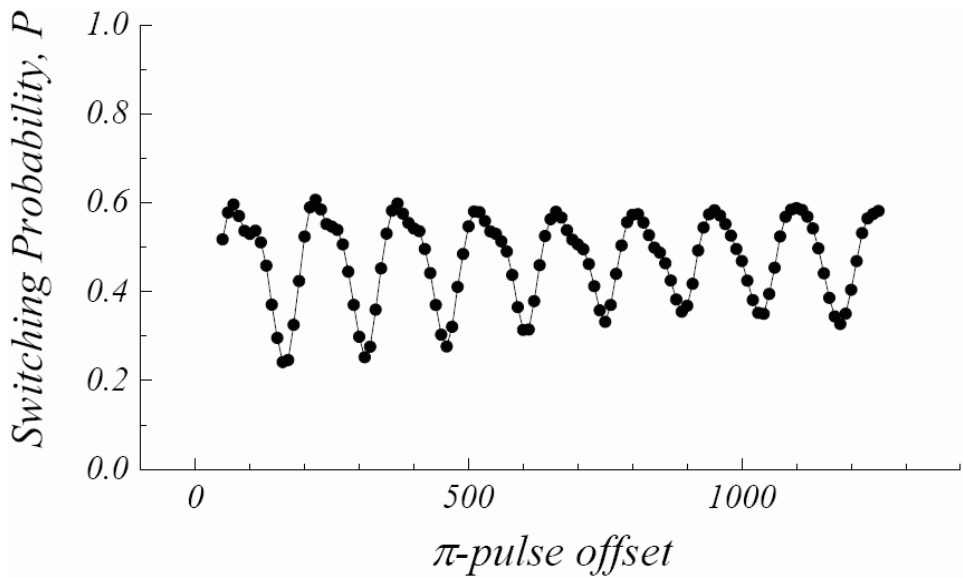}%
\caption{
Spin-echo-type oscillations for three-junction loop. Parameters are $%
M_{dc}=0.52$, $\protect\beta _{L}=0.09559$, $\protect\kappa =0.68$, $\protect%
\alpha =1.5\cdot 10^{-4}$, $\protect\omega _{s}=0.419787$, $\protect%
\varepsilon _{s}=1.82\cdot 10^{-3}$, $\protect\varepsilon _{p}=0.0195$, and $%
\Theta (T)=3\cdot 10^{-3}$. Each dot represents $22000$ escape events.}%
\label{Fig.6}%
\end{center}
\end{figure}

In previous work, and here in Fig.~5b, we have considered
Rabi-type oscillations using the resonant frequency as the driving
frequency: $\omega _{s}=\omega _{l}\sim \omega _{r}$. However, this is not
necessarily the case for all experiments. In Fig.~7 we
show the effect of a lower-than-resonance ($\omega _{s}<\omega _{l}$)
driving signal on Rabi-type frequency. One important aspect of these plots
is that low-amplitude signals result in modulation frequencies which do not
lie on the "main sequence" curve to which higher-amplitude responses
conform. We assert that the reason for this is found in the multi-valued
nature of the amplitude (and energy) response for off-resonant driving as
shown in Ref.~[27], Fig.~1. Also notable is the fact that this effect (an
analytical result for the single junction) is also seen in the
three-junction loop.
\begin{figure}
[pt]
\begin{center}
\includegraphics[
trim=0.000000in 0.000000in 0.000000in 0.000000in,
height=3.00000in,
width=5.000000in
]%
{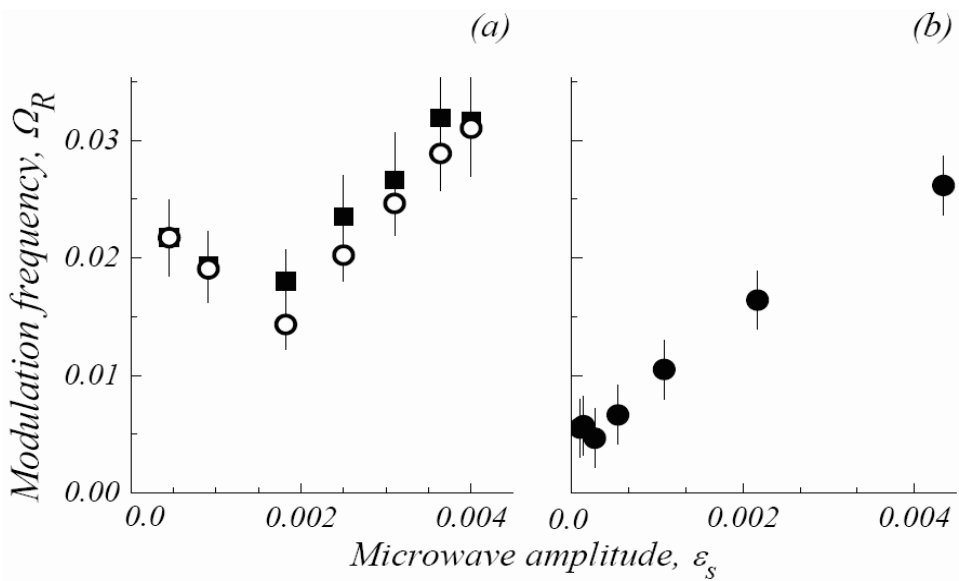}%
\caption{
Simulations
if off-resonance Rabi-type oscillation frequencies. Panel~(a) depicts the
off-resonance modes for the three-junction loop. Parameters were: $%
M_{dc}=0.52$, $\protect\beta _{L}=0.09559$, $\protect\kappa =0.68$, $\protect%
\omega _{s}=0.443$, $\protect\varepsilon _{s}=1.82\cdot 10^{-3}$, $\protect%
\varepsilon _{p}=0.063$, and $\Theta =3\cdot 10^{-3}$. Two values of
characteristic damping are given. The filled squares correspond to $\protect%
\alpha =1.5\cdot 10^{-4}$, empty circles:$\protect\alpha =0$. The resonance
frequency, $\protect\omega _{r}$ (as shown in Fig.~3d)
is $0.45629$. Panel~(b) shows the off-resonance modes for the
single-junction circuit. Parameters were $\protect\alpha \sim 1.5\cdot
10^{-4}$, $\protect\eta =\protect\sqrt{1-\protect\omega _{l}^{4}}$, $\protect%
\omega _{s}=0.99\protect\omega _{l}=0.646188$, and $\Theta =2\cdot 10^{-4}$.}%
\label{Fig.7}%
\end{center}
\end{figure}

\section{Conclusions}

The analysis we have developed reducing the three-junction loop to a single
degree of freedom provides a significant reduction in the complexity of the
system. The resulting equation provides direct agreement with our
simulations when identifying the system resonances from experimentally
relevant switching experiments.

Our studies of off-resonant driving of Rabi-type oscillations reinforce
earlier work concerning multi-valued functions in signal amplitude. Although
no distinct conclusions can be made with regard to a connection between
off-resonant driving signals and linear relationships between Rabi-type
frequency and signal amplitude, we have shown evidence that driving the
system near (but not exactly at) resonance does extract nearly linear
behavior in the response for larger amplitudes. We emphasize that the simple
model we are using may not completely account for all the details of the
observed phenomena.

Consistently with our previous work, we have used the classical RCSJ
phenomenology to investigate various phenomena heretofore attributed to
macroscopic quantum tunneling and \ we have shown very good agreement with
experimental results so far reported. We have shown that, with regard to
Rabi-type oscillations, Ramsey-type fringes, and spin-echo-type
oscillations, the three-junction loop differs little, qualitatively, from
the single-junction circuit. This is easily attributed to the fact that the
important dynamics of these systems is governed by resonant excitations in
anharmonic potential wells followed by a read-out perturbation which makes
the system escape from the well. Thus, our developed analysis and intuition
about the system response to applied microwave pulses translates easily from
a simple single junction system to other systems with similar overall
resonant excitations and escape from anharmonic wells. The main difference
between the single junction system and other systems, such as the three loop
system studied here, is that the experimental reality may be different.
However, given the relative insensitivity to noise inherent in the
three-junction loop, it is understandable that several research groups work
with flux qubits rather than single junctions (biased by dc bias current
supplies).

The work herein presented was developed on the background of the PhD thesis%
\cite{Jeff_Thesis} of one of the authors (JEM) which was submitted to the
University of California at Davis in the fall of 2006. In the present paper
we have chosen to keep the same terminolology (Rabi-type oscillations,
Ramsey-type fringes etc) introduced in the thesis and in the papers which
inspired it \cite%
{jensen:multi-peak_1,jensen:multi-peak_2,jensen:multi-peak_LJJ,cirillo95}.
Other authors have recently reported on RCSJ-based analysis of the
three-junction Josephson system presenting their own terms and analysis\cite%
{nori08}: in these papers, the pumping by an ac drive was used in order to
generate low frequency modulations of phase and energy and the conclusions
confirm a nonlinear phenomenon reported previously in simulations and
experiments\cite%
{jensen:multi-peak_1,jensen:multi-peak_2,jensen:multi-peak_LJJ,cirillo95},
namely that it is possible to phase-lock Josephson systems by superharmonic
pumping. The quantum effects on the other hand should not be sensitive to
the superharmonic drive and therefore this difference could constitute a
discriminant between classical and quantum effects. Two of us (MC and NGJ)
were co-authors in publications\cite{jensen:multi-peak_1,jensen:multi-peak_2}
in which the superharmonic pumping of phase qubit was observed over a broad
temperature range: we recall indeed that nonlinear RCSJ dynamics can model a
Josephson system over broad parameter ranges and in particular at very low
temperatures. Recent observations \cite{catastrophe} indicate that this is
indeed is the case down to 10mK where striking experimental observations of
nonlinear effects are observed ; therefore it might not be easy to engineer
a Josephson system on which all the nonlinear effects are purged out. It is
also worth noticing that it is not straightforward to draw conclusions from
experiments with superharmonic pumping of a superconducting circuit due to
the varying responses that both the system and the microwave apparatus may
have at different frequencies.

The possibility that the behavior of a physical system thought to
demonstrate evidence of macroscopic quantum superposition can be interpreted
with a different approach is not a unique characteristic the Josephson
effect. Roughly three decades ago scientists were looking for macroscopic
quantum effects in charge density-wave systems \cite{Bardeen} and it was
found that the reported results could also be explained by alternative and
more "classical" models \cite{Sneddon}. It was earlier speculated that these
kind of arguments would not find space and motivation in Josephson systems,
but while our findings for resonant switching, Rabi-oscillations,
Ramsey-fringes, and spin-echo do not preclude observations of quantum
behavior in Josephson systems, the evidence of recent years has shown that
many observed phenomena in this class of systems can be attributed to the
RCSJ dynamics developed for Josephson variables in a classical limit\cite%
{vanduzer.barone}.

\section{Acknowledgments}

We wish to thank John Clarke, Alexey Ustinov, Paul Reichardt, and Travis
Hime for taking the time to assist us in this endeavor. This work was
supported in part by the UC Davis Center for Digital Security under the
AFOSR grant FA9550-04-1-0171.


\end{document}